\documentclass[english,smallextended]{svjour3}
\usepackage[T1]{fontenc}
\usepackage[latin9]{inputenc}
\usepackage{xcolor}
\usepackage{pdfcolmk}
\usepackage{float}
\usepackage{marvosym}
\usepackage{url}
\usepackage{bm}
\usepackage{amsmath}
\usepackage{amssymb}
\usepackage{graphicx}
\usepackage{setspace}
\usepackage[authoryear]{natbib}
\PassOptionsToPackage{normalem}{ulem}
\usepackage{ulem}
\makeatletter

\providecommand{\tabularnewline}{\\}
\providecolor{lyxadded}{rgb}{0,0,1}
\providecolor{lyxdeleted}{rgb}{1,0,0}

\usepackage{babel}
\usepackage{bm}\usepackage{marvosym}

\journalname{Brain Topography}

\makeatother

\usepackage{babel}
\begin{document}

\title{The joint use of the tangential electric field and surface Laplacian
in EEG classification}

\author{C. G. Carvalhaes \and J. Acacio de Barros \and M. Perreau-Guimar\~aes
\and P. Suppes}

\institute{C. G. Carvalhaes ({\large \Letter{}})\and M. Perreau-Guimar\~aes \and
P. Suppes \at Suppes Brain Lab, 220 Panama St, Ventura Hall, Stanford,
CA 94305-4115\\
 \email{claudioc@stanford.edu, mpguimaraes@stanford.edu, psuppes@stanford.edu}\and
J. Acacio de Barros \at Liberal Studies, San Francisco State University,
1600 Holloway Ave, San Francisco, CA 94312\\
 \email{barros@sfsu.edu}}
\maketitle
\begin{abstract}
We investigate the joint use of the tangential electric field and
the surface Laplacian derivation as a method to improve the classification
of EEG signals. We considered five classification tasks to test the
validity of such approach. In all five tasks, the joint use of the
components of the tangential electric field and the surface Laplacian
outperformed the scalar potential. The smallest effect occurred in
the classification of a mental task, wherein the average classification
rate was improved by 0.5 standard deviations. The largest effect was
obtained in the classification of visual stimuli and corresponded
to an improvement of 2.1 standard deviations.\\
\end{abstract}

\section*{Introduction}

Advanced techniques of analysis and interpretation of the EEG signals
have grown substantially over the years, with special attention directed
to the problems of low spatial resolution and the choice of physical
reference. Performing the surface Laplacian differentiation of scalp
potentials has proved to be an efficient method to address both issues
\citep{Hjorth1975,Perrin1989,He1992,He1993,Babiloni1995,Yao2002,Nunez2006}.
The surface Laplacian operation is reference-free and many studies
have suggested that it provides a more accurate representation of
dura-surface potentials than conventional topography \citep{Nunez1991,Nunez1991a,Nunez1994,Nunez1994a,Srinivasan1996,Nunez2006}.
Motivated by this observation, numerous experimental studies of both
clinical and theoretical interest have been successfully conducted,
such as those of \citet{Babiloni1999,Babiloni2000a,Babiloni2001,Babiloni2002,Chen2005,Besio2006,Kayser2006a,Kayser2006b,Koka2007,Bai2008,Besio2009}.

In physical terms, the surface Laplacian of the scalp potential is
a measure of local effects of geometry and boundary conditions on
the normal component of the underlying current density. This follows
directly from the quasistatic continuity of the current density, which
implies that any change in flux normal to a surface causes a lateral
divergence of flow lines, which may be expressed mathematically as
the surface Laplacian of the surface potential. This relationship
between the surface Laplacian of the potential and the normal component
of the current density results in a spatial filtering property that
is responsible for the majority of practical applications of the Laplacian
technique. But unless reliable information is available about the
physical process underlying the EEG, relying exclusively on the behavior
of the normal component of the current density may imply the neglect
of potentially valuable information encoded in other spatial components.
This observation was a compelling reason to undertake the present
work. Thus, in our approach we jointly consider the surface Laplacian
of the scalp potential and the tangential components of the scalp
electric field to classify EEG signals.

The rationale for this combination is that the electric field is also
locally related to the current density by Ohm's law. Each spatial
component of the electric field expresses the (negative) rate of change
of the scalar potential in that direction, but because the EEG is
only recorded along the scalp, we cannot estimate the field component
normal to the scalp surface directly from the data. The use of the
surface Laplacian to represent this spatial component is not new in
the literature. For instance, \citet{He1995} discussed the physical
existence of the normal component of the electric field just out of
the body surface and used its analytic relationship with the surface
Laplacian of the potential to construct a surface-charge model to
represent bioelectrical sources inside the body. In the Appendix,
we use similar considerations to explain the connection between these
quantities at electrode sites on a spherical scalp model.

All computations in our work were performed by means of regularized
splines on the sphere. We evaluated the practical effect of the joint
approach on five classification tasks, derived from experiments on
language, visual stimuli, and a mental task. The results in terms
of effect sizes showed an optimistic prospect for further developments
and applications.

\section*{Methods}

\subsection*{Experimental Procedures}

All data used in our study were previously obtained in our laboratory
as part of experiments on language, vision, and imagination. We label
such experiments as Exp.\,I, Exp.\,II, and Exp.\,III, and the subjects
who took part in them as S1, S2, S3, $\cdots$, but S1 of Exp.\,I
was not necessarily the same as S1 of Exp.\,II, and so on. Exp.\,I
encompassed three distinct classification tasks and Exp.\,II and
III one classification task each.

\subsubsection*{Exp.\,I: 32 consonant-vowel syllables}

This experiment emerged from Rui Wang's doctoral work \citep{Wang2011}
and is described in detail in \citet{Wang2012}, to which the reader
is referred for further details. The focus was on the identification
of brain patterns related to listening to a set of English phonemes
having traditional phonological features of consonants (voicing, continuant,
and place of articulation) and vowels (height and backness). The stimuli
consisted of the sounds of 32 phonemes (8 consonants $\times$ 4 vowels)
formed from pairwise combinations of the consonants /p/, /t/, /b/,
/g/, /f/, /s/, /v/, /z/, and the vowels /i/ (as in m\textbf{ee}t),
/�/ (c\textbf{a}t), /u/ (s\textbf{oo}n), and /a/ (sp\textbf{a}). These
vowels were selected also for being maximally separated in the American-English
vowel space, which presumably facilitates classification. All phonemes
were uttered by a male native speaker of English and recorded in audio
files at 44.1\,kHz sampling rate. Each syllable was repeated 7 times
to produce a variation of pronunciation as commonly occurs in human
languages. This resulted in a total of $7\times32=224$ audio files
for presentation.

Four adult subjects (S1-S4), 1 male, participated in this experiment,
all reporting no history of hearing problems. The auditory stimuli
were presented to participants in random order via a stereo computer
speaker. Each participant was instructed to listen carefully to the
sounds and try to comprehend them, but no response was required. The
stimulus presentations were grouped into multiple sessions of 896
trials (4 repetitions of 224 sounds at random). The trial length,
measured from the onset of one stimulus to the onset of the next,
had 1,000\,ms duration, so that each session lasted approximately
15 min. There was a short break after each block of 56 trials, and
the participant could control the length of the break by pressing
the spacebar. The number of trials collected from the participants
were: 7,168 (S1), 3,584 (S2), 6,272 (S3), and 4,480 (S4).

\begin{figure}[H]
\noindent \includegraphics[width=6cm]{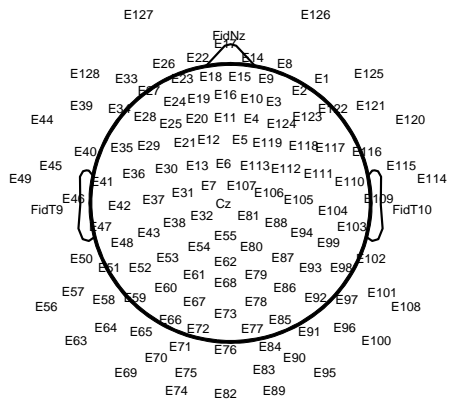}\caption{Electrode montage used in Exp.\,I.\label{fig:Electrode-Positions-Exp1}}
\end{figure}

EEG signals were recorded at 1,000\,Hz sampling rate, using a 128-channel
Geodesic Sensor Net (Figure\,\ref{fig:Electrode-Positions-Exp1})
on EGI's Geodesic EEG system 300. There were 124 monopolar channels
with a common reference Cz and 2 bipolar reference channels for eye
movements. 

\subsubsection*{Exp.\,II: 9 shape-color images}

In this experiment researchers from our Lab investigated
the brainwave representation of nine two-dimensional images, formed
by pairwise combinations of three geometric shapes (\emph{circle},
\emph{square}, \emph{and triangle}) and three colors (\emph{red},
\emph{green}, and \emph{blue}). These images were presented to participants
on a 17-inch LED computer screen using the commercial software Presentation.
All shapes had approximately the same area of 100\,cm$^{2}$, and
the physical luminosity was adjusted for each object and color to
appear the same at 60\,cm from the screen, but no attempt was made
to adjust images to each participant, such that the subjective perception
of luminosity was the same for all colors and each participant. The
distance from the participant' eyes to the screen was approximately
60\,cm and the visual angle was 2.3\textendash{}3.3$^{\circ}$. Each
presentation lasted 300\,ms and was followed immediately by an interval
of 700\,ms, during which a fixation cross (`$+$') was shown at the
center of a blank screen. A stimulus appeared randomly and with equal
probability every 1,000\,ms.

Seven adults (S1-S7), 3 female, agreed to participate in this experiment,
all having normal or corrected to normal vision. Participants were
instructed to remain relaxed and motionless, and to keep eyes fixed
at the center of the screen during presentations. They were seated
comfortably on a chair in a dimly lit sound-attenuated booth and responded
to 2,700 trials time-locked to stimulus presentations. The presentations
were divided in blocks of 20 trials and the participant could control
the duration of the breaks via the spacebar. Halfway through the experiment
a modified break message was displayed informing the participant that
the experiment had passed its halfway point. 
\begin{figure}[H]
\noindent \includegraphics[width=5cm]{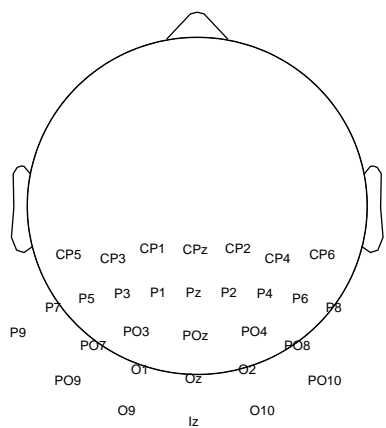}\caption{Electrode montage used in Experiment II.\label{fig:Electrode-Positions-Exp2}}
\end{figure}

\subsubsection*{Exp.\,III: 2-class imagery task}

The third experiment was previously described in \citet{Carvalhaes2009}
and \citet{Carvalhaes2011}. Eleven participants (S1-S11) were randomly
presented on every other trial either a visual ``stop'' sign, flashed
on a 17-inch LED computer screen, or the sound of the English word
``go'', via computer speaker. The ``go'' sound was uttered by
a male native speaker of English and recorded in an audio-isolated
cabin using a professional microphone interfaced with a computer via
a Sound Blaster II (Creative Labs) sound card at 44.1\,kHz sampling
rate (24~bits). Stimuli were delivered using the Presentation software.
The ``go'' sound was delivery via high fidelity PC speakers at the
level of normal conversation. Each stimulus presentation lasted 300\,ms,
and was followed by a period of 700\,ms of blank screen. Immediately
after this period a fixation cross (`$+$') was shown at the center
of the screen for 300\,ms.

Eleven subjects participated in this experiment, all adults reporting
normal vision and normal hearing. They were comfortably seated in
a chair at a distance of approximately 60\,cm from a computer screen
and 100\,cm from the speaker. For one group (S1-S7) the participants
were instructed to form a vivid mental image of the stimulus previously
presented, for another group (S8-S11) they were asked to form a mental
image of the alternative stimulus, i.e., if the last stimulus was
the ``stop'' sign, then they should imagine the ``go'' sound,
and vice versa. Participants' imagining was followed by another 700\,ms
of blank screen, after which the trial ended. A single session of
600 trials was recorded for each participant. The session was divided
into thirty 20-trial blocks, with regular breaks controlled by participants
via the spacebar. Each trial lasted 2,000\,ms, but only the last
1,000\,ms of each trial corresponding to the imagination task was
used for our analysis. 
\begin{figure}[H]
\noindent \includegraphics[width=6cm]{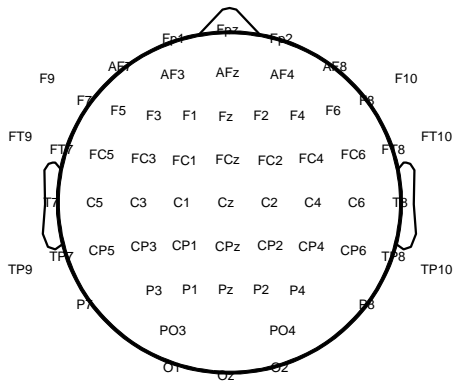}\caption{Electrode montage used in Experiment III.\label{fig:Electrode-Positions-Exp3}}
\end{figure}

\subsection*{Data collection and preprocessing}

Data collection started after participants were given the opportunity
to practice the required tasks. The recording apparatus changed from
one experiment to another. Exp.\,I was carried out using EGI's Geodesic
EEG system with 128 monopolar channels referenced to the vertex electrode
(Cz) and with a ground electrode placed on the forehead (also for
Exp II and III). The electrode locations for this experiment are illustrated
in Figure\,\ref{fig:Electrode-Positions-Exp1}. In Exp.\,II signals
were recorded using a 32-channel NeuroScan system with linked earlobe
reference (Ag\textendash{}AgCl electrodes). Due to the low number
of channels available on this device -- and in view of the need for
a reasonable density of electrodes to accurately estimate the electric
field in the region of interest V1 \citep{Mikkulainen2005} -- the
measurement electrodes were all placed in the back part of the head,
as depicted in Figure\,\ref{fig:Electrode-Positions-Exp2}. We remark
that there was no particular reason for choosing P9 instead of P10
in this montage. The electrode distribution was asymmetric and P10
was not included as well just because of the small number of channels
that were available to perform this experiment. Exp.\,III used a
64-channel Neuroscan system, following the 5\% system of \citet{Oostenveld2001},
but not including electrodes Nz, AF1, AF2, AF5, AF6, T9, T10, P9,
P5, P6, P9, P10, PO, or I; the reference being as in Exp.\,II. Figure\,\ref{fig:Electrode-Positions-Exp3}
shows the electrode distribution for this experiment.

The signals were passed through a band-pass filter in the range 0.1-300\,Hz
plus a 60\,Hz notch filter, and digital conversion was performed
at a 1\,kHz sampling rate. To reduce features, we carried out offline
decimation at 16:1 ratio, thus setting the Nyquist frequency at 31.25\,Hz.
Additionally, we removed unwanted low-frequency components by applying
a high-pass filter of 1\,Hz. Finally, we mathematically referenced
the decimated signals to the average reference voltage to reduce biases
in the analysis of the potential distribution \citep{Bertrand1985,Murray2008}.
This step had no effect on the tangential field and the surface Laplacian
derivation, for they are reference-free quantities \citep{He1993,He1995,Nunez2006}.

The signals were visually inspected, but no trial was removed. Thus,
robustness to outliers and artifacts was also tested in the classification.
In order to enhance the signal-to-noise ratio, we averaged same-class
trials over small groups of trials before classification. We fixed
the number of samples per average trial according to the amount of
classes and the total number of trials available in the experiment.
With this constraint in mind, the number of samples per average trial
was set to: 12 (Exp.\,I, 8 initial consonants); 5 (Exp.\,I, 32 syllables);
20 (Exp.\,I, 4 vowels); 5 (Exp.\,II); and 5 (Exp.\,III).

\subsection*{Numerical procedure}

For convenience, we adopted spherical coordinates $(r,\,\theta,\,\varphi)$,
where $r$ stands for radial distance and $\theta$ and $\varphi$
are the angular coordinates, with $\theta$ increasing down from the
vertex and $\varphi$ increasing counterclockwise from the nasion.
The scalar potential, the tangential components of the electric field,
and the surface Laplacian of the potential were denoted by $\Phi_{s}^{\mathrm{scalp}}$,
$E_{\theta}^{\mathrm{scalp}}$, $E_{\varphi}^{\mathrm{scalp}}$, and
$\nabla_{s}^{2}\Phi_{s}^{\mathrm{scalp}}$. The mathematical expressions
for these quantities are shown in the Supplementary Material in terms
of partial derivatives of $\Phi_{s}^{\mathrm{scalp}}$. To obtain
these quantities we fitted $\Phi_{s}^{\mathrm{scalp}}$ with a spline
interpolant and then applied the partial derivatives analytically
to the interpolant. This computation was carried out using $\lambda$-correction
to attenuate the effect of spatial noises on the estimates \citep{Wahba1990,Babiloni1995}.

Using splines we can calculate partial derivatives at a very low computational
cost. Assume an instantaneous distribution of scalp potentials $\{V_{1},\cdots,V_{N}\}$,
sampled at electrode locations $\mathbf{r}_{1},\cdots,\mathbf{r}_{N}$
at a time $t$. The spline interpolant that fits or smooths this distribution
is defined by 
\begin{equation}
f_{\lambda}(\mathbf{r})=\sum_{j=1}^{N}c_{j}\,\left\Vert \mathbf{r}-\mathbf{r}_{j}\right\Vert ^{2m-3}+\sum_{\ell=1}^{M}d_{\ell}\,\phi_{\ell}(\mathbf{r}),\label{eq:interpolant}
\end{equation}
where $m$ is an integer greater than 2, $M=\binom{m+2}{3}$ is subject
to $M<N$, $\phi_{1},\cdots,\phi_{M}$ are linearly-independent polynomials
in $\mathbb{R}^{3}$ of degree less than $m$, and $c_{j}$ and $d_{j}$
are data-dependent parameters. In order to avoid the magnification
of high-frequency spatial noises, we introduce a regularization parameter,
$\lambda$, such that \citep{Wahba1990,Babiloni1995} 
\begin{equation}
\left(\begin{array}{cc}
\mathbf{K}+N\lambda\mathbf{I} & \mathbf{T}\\
\mathbf{T}^{\prime} & 0
\end{array}\right)\left(\begin{array}{c}
\mathbf{c}\\
\mathbf{d}
\end{array}\right)=\left(\begin{array}{c}
\mathbf{v}\\
\mathbf{0}
\end{array}\right),\label{eq:interpolation conditions}
\end{equation}
where $(\mathbf{K})_{ij}=\left\Vert \mathbf{r}_{i}-\mathbf{r}_{j}\right\Vert ^{2m-3}$,
$(\mathbf{T})_{ij}=\phi_{j}(\mathbf{r}_{i})$, $\mathbf{c}=(c_{1},\cdots,c_{N})^{\prime}$,
$\mathbf{d}=(d_{1},\cdots,d_{M})^{\prime}$, and $\mathbf{v}=(V_{1},\cdots,V_{N})^{\prime}$.
As explained in \citet{Carvalhaes2011} and \citet{Carvalhaes2013},
the system \eqref{eq:interpolation conditions} is singular on a spherical
surface, so that we can not obtain $\mathbf{c}$ and $\mathbf{d}$
by just inverting this system. Instead, following \citet{Carvalhaes2011}
we factorize $\mathbf{T}$ as 
\begin{equation}
\mathbf{T}=\left(\mathbf{Q_{1}}\,,\,\mathbf{Q_{2}}\right)\,\left(\begin{array}{c}
\mathbf{R}\\
O
\end{array}\right),
\end{equation}
where $\mathbf{Q}_{1}\in\mathbb{R}^{N\times M}$ and $\mathbf{Q}_{2}\in\mathbb{R}^{N\times(N-M)}$
are orthonormal and $\mathbf{R}\in\mathbb{R}^{M\times M}$ is upper
triangular, and introduce the auxiliary matrices\begin{subequations}
\begin{gather}
\mathbf{C}_{\lambda}=\mathbf{Q}_{2}\left[\mathbf{Q}_{2}^{\prime}\left(\mathbf{K}+N\lambda\mathbf{I}\right)\mathbf{Q}_{2}\right]^{-1}\mathbf{Q}_{2}^{\prime},\\
\mathbf{D}_{\lambda}=\mathbf{R}^{+}\mathbf{Q}_{1}^{\prime}\left(\mathbf{1}-\mathbf{K}\mathbf{C}_{\lambda}-N\lambda\mathbf{C}_{\lambda}\right),
\end{gather}
\end{subequations}where $\mathbf{R}^{+}$ is the pseudo-inverse of
$\mathbf{R}$.

Let $\mathbf{e}_{\theta,\lambda}$, $\mathbf{e}_{\varphi,\lambda}$,
and $\mathbf{l}_{\lambda}$ be $N$-dimensional vectors giving $E_{\theta}^{\mathrm{scalp}}$,
$E_{\varphi}^{\mathrm{scalp}}$, and $\nabla_{s}^{2}\Phi_{s}^{\mathrm{scalp}}$
at the electrode coordinates. These vectors can be obtained by linearly
transforming the potential $\mathbf{v}$ as\begin{subequations}\label{eq:diff_matrices}
\begin{align}
\mathbf{e}_{\theta,\lambda} & =\big(\mathbf{K}_{\theta}\mathbf{C}_{\lambda}+\mathbf{T}_{\theta}\mathbf{D}_{\lambda}\big)\mathbf{v}=\mathbf{E}_{\theta,\lambda}\mathbf{v},\label{eq:diff_matrices:e_theta}\\
\mathbf{e}_{\varphi,\lambda} & =\big(\mathbf{K}_{\varphi}\mathbf{C}_{\lambda}+\mathbf{T}_{\varphi}\mathbf{D}_{\lambda}\big)\mathbf{v}=\mathbf{E}_{\varphi,\lambda}\mathbf{v},\label{eq:diff_matrices:e_phi}\\
\mathbf{l}_{\lambda} & =\big(\tilde{\mathbf{K}}\mathbf{C}_{\lambda}+\tilde{\mathbf{T}}\mathbf{D}_{\lambda}\big)\mathbf{v}=\mathbf{L}_{\lambda}\mathbf{v}.\label{eq:diff_matrices:Laplacian}
\end{align}
\end{subequations}The analytic expressions for the matrices $\mathbf{K}_{\theta}$,
$\mathbf{T}_{\theta}$, $\mathbf{K}_{\varphi}$, $\mathbf{T}_{\varphi}$,
$\tilde{\mathbf{K}}$, and $\tilde{\mathbf{T}}$ are given in the
supplementary material, along with a Matlab code implementation. The
fact that $\mathbf{e}_{\theta,\lambda}$, $\mathbf{e}_{\varphi,\lambda}$,
and $\mathbf{l}_{\lambda}$ are reference free implies that 
\begin{equation}
\mathbf{E}_{\theta,\lambda}\,\mathbf{v}_{\mathrm{ref}}=\mathbf{E}_{\varphi,\lambda}\,\mathbf{v}_{\mathrm{ref}}=\mathbf{L}_{\lambda}\,\mathbf{v}_{\mathrm{ref}}=0,\label{eq:reference_free}
\end{equation}
where $\mathbf{v}_{\mathrm{ref}}=\mbox{const.}\times(1,\cdots,1)^{T}$
is a reference vector. That is, the columns of $\mathbf{E}_{\lambda}^{\theta}$,
$\mathbf{E}_{\lambda}^{\varphi}$, and $\mathbf{L}_{\lambda}$ sum
to zero, regardless of the value of $\lambda$. A skeptical reader
is encouraged to use the Matlab code in the Supplementary Material
to test this property.

\subsection*{Classification procedures}

For statistical comparison, each experiment was classified using the
potential, the surface Laplacian of the potential, the tangential
electric field, and a combination of the last two into a three-dimensional
vector. We carried out the classifications on single channels, using
a 10-fold cross-validation on linear discriminant analysis (LDA) \citep{Parra2008,Suppes2009}.
For this purpose, the data from each channel and waveform were rearranged
in a rectangular matrix, with adjacent rows corresponding to adjacent
trials and adjacent columns corresponding to adjacent time samples.
Matrices representing vector quantities were given by the concatenation
of the individual components.

Preliminary classifications were performed with a small number of
trials, attempting to find a plausible range of values for the $\lambda$
parameter. This parameter regulates the trade-off between minimizing
the squared error of the data fitting and smoothness \citep{Wahba1990},
thus influencing spatial differentiations and the classification rates.
The overall most satisfactory results were obtained in a grid with
50 points, covering the interval $\lambda\in[0.001\,,\,100]$ in logarithm
scale. Hence, for all tasks the classification of each channel was
repeated 50 times, varying $\lambda$ across this interval.

To further improve the classification rates we applied principal component
analysis (PCA) \citep{Jolliffe2005}, but only the $\lambda$-value
yielding the highest classification rate was considered in this step.
The classification of PCA-transformed data began with the classification
of the first principal component, which ordinarily explains most of
the variance of the data. The other components account for residual
variance and were added in order of decreasing variance. For the purpose
of pairwise comparison the same random sequence of trials was used
in the classification of all waveforms.

\subsection*{Statistical analysis}

We used effect sizes and confidence intervals to assess improvements
in classification rates in comparison to the scalar potential. Effect
size is a standard measure that addresses the practical relevance
of differences in paired comparisons. Typically, it is calculated
by dividing the difference between the means of two groups by the
combined (pooled) standard deviation, i.e., 
\begin{equation}
d=\frac{\mu_{A}-\mu_{B}}{s_{\mathrm{pool}}},\label{eq:Effect_Size}
\end{equation}
where $d$ stands for effect size (also known as Cohen's~$d$),
$\mu_{A}$ and $\mu_{B}$ are the mean values of the two groups, and
$s_{\mathrm{pool}}=\sqrt{(s_{A}^{2}+s_{B}^{2})/2}$ is the pooled
standard deviation ($s_{A}^{2}$ and $s_{B}^{2}$ are the respective
variances). We used equation \eqref{eq:Effect_Size} with $B$ standing
for the potential and $A$ standing for the tested waveform.

Intuitively, equation \eqref{eq:Effect_Size} expresses how many standard
deviations separate the performance of two methods; the larger the
effect size, the greater the performance of the tested method. A zero
effect size indicates a failure in rejecting the null hypothesis of
no difference between the methods. In other words, the effect size
has the following practical application: it tells us not only whether
the null hypothesis is being rejected, but also gives us a sense of
the strength of this rejection. In contrast to null-hypothesis significance
testing (often represented by a $p$-value), the effect size is not
particularly sensitive to the sample size, and hence it can be compared
across different studies, even though the number of samples is not
the same.

In order to make our statistical comparison more reliable, we estimated
a confidence limit around each effect size. Namely, the null hypothesis
of no practical effect of the tested waveform was rejected at the
95\% level of significance only if the estimated confidence interval
did not include zero. The confidence interval of $d$ was estimated
by the equation 
\begin{equation}
\mbox{95\%CI}=[d-1.96\times\mbox{SE},d+1.96\times\mbox{SE}],
\end{equation}
where SE is the standard error between the paired rates from $A$
and $B$, given by 
\begin{equation}
\mathrm{SE}=\sqrt{\frac{2(1-r_{AB})}{n_{p}}+\frac{d^{2}}{2(n_{p}-1)}}\label{eq:standard_error}
\end{equation}
where $n_{p}$ is the number of participants in the experiment and
$r_{AB}$ is the correlation coefficient for the paired rates \citep{Becker1988,Nakagawa2007}.
Note that equation \eqref{eq:standard_error} depends on the sample
size $n_{p}$. Large samples yield small errors and, consequently,
narrow confidence intervals. In contrast, small samples provide less
focused estimates of the effect size, but this cannot be mistaken
as evidence for a null effect, as usually occurs when reporting $p$-values.
This remark is particularly important to our study because $n_{p}$
was generally small, thus resulting in large confidence intervals.

\section*{Results}

\subsection*{Classification rates}

Tables\,\ref{tab:Exp1a}-\ref{tab:ExpIII} summarize the classification
outcomes, showing the highest cross-validation rate of each task,
along with the best sensor. The surface Laplacian of the scalp potential
and the tangential electric field are referred to by SL and EF, and
their combination by SL \& EF. Bearing in mind the chance level of
each task, generally the rates were remarkably good. In Table\,\ref{tab:Exp1a}
we show the classification rates for Exp.\,I using the initial consonants
to define the eight classes. The combination of the surface Laplacian
and the tangent electric field not only yielded the highest classification
rate for all participants, but interestingly its best performance
was achieved by locations on the primary auditory cortex A1 \citep{Pickles2012}
for all subjects. Averaged over all participants, this resulted in
a improvement of 10.6\% in comparison with the potential and 4.8\%
in relation to the surface Laplacian of the potential. The tangential
electric field had a similar performance, but with individual classification
rates being slightly smaller for all participants. 
\begin{table}[H]
\noindent \textsf{\caption{\textsf{Highest performance for the classification of the 8 initial
consonants of Exp.\,I.\label{tab:Exp1a}}}
}

\begin{tabular}{lrllrllllcr@{\extracolsep{0pt}.}ll}
\hline 
 & \multicolumn{2}{c}{\textsf{potential}} &  & \multicolumn{2}{c}{\textsf{SL}} &  & \multicolumn{2}{c}{\textsf{EF}} &  & \multicolumn{3}{c}{\textsf{SL \& EF}}\tabularnewline
\hline 
\textsf{\small subject} & \multicolumn{1}{c}{\textsf{\small \%}} & \textsf{\small sensor}  &  & \multicolumn{1}{c}{\textsf{\small \%}} & \textsf{\small sensor} &  & \textsf{\small \%} & \textsf{\small sensor} &  & \multicolumn{2}{c}{\textsf{\small \%}} & \textsf{\small sensor}\tabularnewline
\hline 
\textsf{\small S1} & \textsf{\small 45.2} & \textsf{\small E36} &  & \textsf{\small 57.5} & \textsf{\small E41} &  & \textsf{\small 60.7} & \textsf{\small E40} &  & \textsf{\small 63}&{\small 0} & \textsf{\small E40,E41}\tabularnewline
\textsf{\small S2} & \textsf{\small 37.2} & \textsf{\small E30} &  & \textsf{\small 40.5} & \textsf{\small E12} &  & \textsf{\small 43.4} & \textsf{\small E109} &  & \textsf{\small 44}&\textsf{\small 4} & \textsf{\small E40}\tabularnewline
\textsf{\small S3} & \textsf{\small 27.5} & \textsf{\small E13,E29} &  & \textsf{\small 32.6} & \textsf{\small E28} &  & \textsf{\small 36.0} & \textsf{\small E47} &  & \textsf{\small 38}&\textsf{\small 8} & \textsf{\small E35}\tabularnewline
\textsf{\small S4} & \textsf{\small 31.9} & \textsf{\small E112} &  & \textsf{\small 30.3} & \textsf{\small E20} &  & \textsf{\small 31.9}  & \textsf{\small E122} &  & \textsf{\small 32}&\textsf{\small 4} & \textsf{\small E41}\tabularnewline
\hline 
\textsf{\small Avg.$\pm$std.} & \multicolumn{2}{l}{\textsf{\small 35.9$\pm$7.4}} &  & \multicolumn{2}{l}{\textsf{\small 41.7$\pm$11.9}} &  & \multicolumn{2}{l}{\textsf{\small 44.6$\pm$12.1}} &  & \multicolumn{3}{l}{\textsf{\small 46.5$\pm$12.5}}\tabularnewline
\hline 
\multicolumn{13}{l}{\textsf{\small Number of trials per participant: S1\,-\,600, S2\,-\,304,
S3\,-\,528, and S4\,-\,376.}}\tabularnewline
\multicolumn{13}{l}{\textsf{\small Chance level: 12.5\%.}}\tabularnewline
\end{tabular}
\end{table}

Table\,\ref{tab:Exp1b} shows the highest rate for the classification
of the 32 syllables of Exp.\,I. The number of classes was four times
larger than the number of initial consonants, which resulted in a
reciprocal decrease in classification accuracy. Once again the SL
\& EF provided the best results, except for S4, for which it yielded
the rate 8.1\% vs. 8.2\% from EF. For this subject, the highest rates
of both methods were achieved in the region of the secondary auditory
cortex (A2). 
\begin{table}[H]
\noindent \textsf{\caption{\textsf{Highest performance for the classification of the 32 syllables
of Exp.\,I. \label{tab:Exp1b}}}
}

\begin{tabular}{crlcrlcrlcrl}
\hline 
 & \multicolumn{2}{c}{\textsf{potential}} &  & \multicolumn{2}{c}{\textsf{SL}} &  & \multicolumn{2}{c}{\textsf{EF}} &  & \multicolumn{2}{c}{\textsf{SL \& EF}}\tabularnewline
\hline 
\textsf{\small subject} & \multicolumn{1}{c}{\textsf{\small \%}} & \textsf{\small sensor}  &  & \multicolumn{1}{c}{\textsf{\small \%}} & \textsf{\small sensor} &  & \multicolumn{1}{c}{\textsf{\small \%}} & \textsf{\small sensor} &  & \multicolumn{1}{c}{\textsf{\small \%}} & \textsf{\small sensor}\tabularnewline
\hline 
\textsf{\small S1} & \textsf{\small 13.4} & \textsf{\small E36} &  & \textsf{\small 21.4} & \textsf{\small E41} &  & \textsf{\small 19.6} & \textsf{\small E46} &  & \textsf{\small 23.1} & \textsf{\small E41}\tabularnewline
\textsf{\small S2} & \textsf{\small 9.5} & \textsf{\small E30} &  & \textsf{\small 9.9} & \textsf{\small Cz} &  & \textsf{\small 10.9} & \textsf{\small E35} &  & \textsf{\small 11.0} & \textsf{\small E40}\tabularnewline
\textsf{\small S3} & \textsf{\small 7.8} & \textsf{\small E13} &  & \textsf{\small 8.4} & \textsf{\small E20} &  & \textsf{\small 9.3} & \textsf{\small E97} &  & \textsf{\small 9.4} & \textsf{\small E44,E46}\tabularnewline
\textsf{\small S4} & \textsf{\small 7.9} & \textsf{\small E6,E13} &  & \textsf{\small 7.1} & \textsf{\small E12} &  & \textsf{\small 8.2} & \textsf{\small E116,E122} &  & \textsf{\small 8.1} & \textsf{\small E116,E122}\tabularnewline
\hline 
\textsf{\small Avg.$\pm$std.} & \multicolumn{2}{l}{\textsf{\small 10.0$\pm$2.5}} &  & \multicolumn{2}{l}{\textsf{\small 12.7$\pm$6.3}} &  & \multicolumn{2}{l}{\textsf{\small 12.8$\pm$5.0}} &  & \multicolumn{2}{l}{\textsf{\small 14.0$\pm$6.7}}\tabularnewline
\hline 
\multicolumn{12}{l}{\textsf{\small Number of trials per participant: S1\,-\,1440, S2\,-\,736,
S3\,-\,1280, and S4\,-\,897.}}\tabularnewline
\multicolumn{12}{l}{\textsf{\small Chance level: 3.1\%.}}\tabularnewline
\end{tabular}
\end{table}

Table\,\ref{tab:Exp1c} summarizes the classification result for
the 4 vowels of Exp.\,I. The rates were significantly above the chance
probability (25\%), but the highest rate (46.9\%, S1) was significantly
smaller than in the classification of the initial consonants (63\%,
S1), which had twice as many classes. Furthermore, improvements in
comparison with the scalar potential were not as large in average
as in the previous two cases.

\begin{table}[H]
\noindent \textsf{\caption{\textsf{Highest performance for the classification of the 4 vowels
of Exp.\,I.\label{tab:Exp1c}}}
}

\begin{tabular}{crlcrlcrlcrl}
\hline 
 & \multicolumn{2}{c}{\textsf{potential}} &  & \multicolumn{2}{c}{\textsf{SL}} &  & \multicolumn{2}{c}{\textsf{EF}} &  & \multicolumn{2}{c}{\textsf{SL \& EF}}\tabularnewline
\hline 
\textsf{\small subject} & \multicolumn{1}{c}{\textsf{\small \%}} & \textsf{\small sensor}  &  & \multicolumn{1}{c}{\textsf{\small \%}} & \textsf{\small sensor} &  & \multicolumn{1}{c}{\textsf{\small \%}} & \textsf{\small sensor} &  & \multicolumn{1}{c}{\textsf{\small \%}} & \textsf{\small sensor}\tabularnewline
\hline 
\textsf{\small S1} & \textsf{\small 40.8} & \textsf{\small E37} &  & \textsf{\small 46.4} & \textsf{\small E42} &  & \textsf{\small 41.1} & \textsf{\small E40} &  & \textsf{\small 46.9} & \textsf{\small E41}\tabularnewline
\textsf{\small S2} & \textsf{\small 38.3} & \textsf{\small E52} &  & \textsf{\small 38.3} & \textsf{\small E80} &  & \textsf{\small 43.9} & \textsf{\small E56} &  & \textsf{\small 43.3} & \textsf{\small E56}\tabularnewline
\textsf{\small S3} & \textsf{\small 37.0} & \textsf{\small E6} &  & \textsf{\small 39.6} & \textsf{\small E127} &  & \textsf{\small 38.3} & \textsf{\small E52} &  & \textsf{\small 38.6} & \textsf{\small E111}\tabularnewline
\textsf{\small S4} & \textsf{\small 39.6} & \textsf{\small E76} &  & \textsf{\small 39.6} & \textsf{\small E76} &  & \textsf{\small 44.4} & \textsf{\small E97} &  & \textsf{\small 41.8} & \textsf{\small E105}\tabularnewline
\hline 
\textsf{\small Avg.$\pm$std.} & \multicolumn{2}{l}{\textsf{\small 39.0$\pm$1.6}} &  & \multicolumn{2}{l}{\textsf{\small 41.6$\pm$3.5}} &  & \multicolumn{2}{l}{\textsf{\small 41.4$\pm$2.5}} &  & \multicolumn{2}{l}{\textsf{\small 42.8$\pm$3.4}}\tabularnewline
\hline 
\multicolumn{12}{l}{\textsf{\small Number of trials per participant: S1\,-\,360, S2\,-\,180,
S3\,-\,316, S4\,-\,225.}}\tabularnewline
\multicolumn{12}{l}{\textsf{\small Chance level: 25.0\%.}}\tabularnewline
\end{tabular}
\end{table}

Table\,\ref{tab:ExpII} shows the classification rates of Exp.\,II.
The lowest classification rate was 52.1\% for subject S2 using the
surface Laplacian of the scalp potential. The highest rate 86.9\%
occurred for subject S1 with SL \& EF. Overall the results were remarkably
good taking into account the chance probability of 11.1\%. In average, the classification rates obtained with EF and SL \& EF were much higher than those obtained with the potential and SL. The SL performed similarly to the potential in average (63.0\% vs. 61.9\%) and rendered the highest standard deviation among the four methods.

\begin{table}[H]
\noindent \textsf{\caption{\textsf{Highest performance for the classification of the 9 images
of Exp.\,II. \label{tab:ExpII}}}
}

\begin{tabular}{r||@{\extracolsep{0pt}.}lrlcrlcrlcrl}
\hline 
\multicolumn{2}{c}{} & \multicolumn{2}{c}{\textsf{potential}} &  & \multicolumn{2}{c}{\textsf{SL}} &  & \multicolumn{2}{c}{\textsf{EF}} &  & \multicolumn{2}{c}{\textsf{SL \& EF}}\tabularnewline
\cline{3-13} 
\multicolumn{2}{c}{\textsf{\small subject}} & \multicolumn{1}{c}{\textsf{\small \%}} & \textsf{\small sensor}  &  & \multicolumn{1}{c}{\textsf{\small \%}} & \textsf{\small sensor} &  & \multicolumn{1}{c}{\textsf{\small \%}} & \textsf{\small sensor} &  & \multicolumn{1}{c}{\textsf{\small \%}} & \textsf{\small sensor}\tabularnewline
\hline 
\multicolumn{2}{c}{\textsf{\small S1}} & \textsf{\small 72.7}{\small{} } & \textsf{\small PO8} &  & \textsf{\small 81.6}{\small{} } & \textsf{\small PO8} &  & \textsf{\small 83.2}{\small{} } & \textsf{\small PO4} &  & \textsf{\small 86.9}{\small{} } & \textsf{\small PO8}\tabularnewline
\multicolumn{2}{c}{\textsf{\small ~S2$^{\mathsf{a}}$}} & \textsf{\small 58.2}{\small{} } & \textsf{\small O9} &  & \textsf{\small 52.1}{\small{} } & \textsf{\small O2} &  & \textsf{\small 68.0}{\small{} } & \textsf{\small PO4} &  & \textsf{\small 71.1}{\small{} } & \textsf{\small O2}\tabularnewline
\multicolumn{2}{c}{\textsf{\small S3}} & \textsf{\small 61.7}{\small{} } & \textsf{\small POz} &  & \textsf{\small 60.6}{\small{} } & \textsf{\small P3} &  & \textsf{\small 62.2}{\small{} } & \textsf{\small CP2} &  & \textsf{\small 70.7}{\small{} } & \textsf{\small P3,P4}\tabularnewline
\multicolumn{2}{c}{\textsf{\small ~S4$^{\mathsf{b}}$}} & \textsf{\small 59.5}{\small{} } & \textsf{\small POz} &  & \textsf{\small 63.9}{\small{} } & \textsf{\small PO8} &  & \textsf{\small 68.1}{\small{} } & \textsf{\small O2} &  & \textsf{\small 75.1}{\small{} } & \textsf{\small PO8}\tabularnewline
\multicolumn{2}{c}{\textsf{\small S5}} & \textsf{\small 66.1}{\small{} } & \textsf{\small PO3} &  & \textsf{\small 70.5}{\small{} } & \textsf{\small POz} &  & \textsf{\small 76.2}{\small{} } & \textsf{\small P1} &  & \textsf{\small 81.8}{\small{} } & \textsf{\small POz}\tabularnewline
\multicolumn{2}{c}{\textsf{\small ~S6$^{\mathsf{a}}$}} & \textsf{\small 60.3}{\small{} } & \textsf{\small O1} &  & \textsf{\small 55.9}{\small{} } & \textsf{\small O2} &  & \textsf{\small 65.8}{\small{} } & \textsf{\small CP4} &  & \textsf{\small 70.0}{\small{} } & \textsf{\small Oz}\tabularnewline
\multicolumn{2}{c}{\textsf{\small ~S7$^{\mathsf{a}}$}} & \textsf{\small 54.8}{\small{} } & \textsf{\small Iz} &  & \textsf{\small 56.6}{\small{} } & \textsf{\small P2} &  & \textsf{\small 67.5}{\small{} } & \textsf{\small P4} &  & \textsf{\small 70.6}{\small{} } & \textsf{\small P4}\tabularnewline
\hline 
\multicolumn{2}{r}{\textsf{\small Avg}{\small{} }\textsf{\small $\pm$std.}} & \multicolumn{2}{l}{\textsf{\small 61.9$\pm$5.9}} &  & \multicolumn{2}{l}{\textsf{\small 63.0$\pm$10.2}} &  & \multicolumn{2}{l}{\textsf{\small 70.2$\pm$7.1}} &  & \multicolumn{2}{l}{\textsf{\small 75.2$\pm$6.7}}\tabularnewline
\hline 
\multicolumn{13}{l}{\textsf{\small $^{\mathsf{a}}$Channels CP6 and PO9 were off. $^{\mathsf{b}}$Channel
CP6 was off. }}\tabularnewline
\multicolumn{13}{l}{\textsf{\small Number of trials per participant: 543 . Chance level:
11.1\%.}}\tabularnewline
\end{tabular}
\end{table}

The classification rates for the trials of the mental task of Exp.\,III
are shown in Table\,\ref{tab:ExpIII}. These rates were higher than
those shown in \citet{Carvalhaes2011} because of the averaging of
trials to reduce temporal noise. Here, most of the best predictions
were achieved by the tangential electric field (S2, S3, S4, S7, S9,
S10, S11) rather than by SL \& EF, which yielded the highest rate
for 5 participants (S3, S6, S8, S9, S11). The SL was the most accurate
method for participants S1 and S5. 
\begin{table}[H]
\noindent \textsf{\caption{\textsf{Highest performance for the classification of the mental task
of Exp.\,III.\label{tab:ExpIII}}}
}

\begin{tabular}{crllrllrllrl}
\hline 
\multicolumn{1}{c}{} & \multicolumn{2}{c}{\textsf{potential}} &  & \multicolumn{2}{c}{\textsf{SL}} &  & \multicolumn{2}{c}{\textsf{EF}} &  & \multicolumn{2}{c}{\textsf{SL \& EF}}\tabularnewline
\hline 
\textsf{\small subject} & \multicolumn{1}{c}{\textsf{\small \%}} & \textsf{\small sensor}  &  & \multicolumn{1}{c}{\textsf{\small \%}} & \textsf{\small sensor} &  & \multicolumn{1}{c}{\textsf{\small \%}} & \textsf{\small sensor} &  & \multicolumn{1}{c}{\textsf{\small \%}} & \textsf{\small sensor}\tabularnewline
\hline 
\textsf{\small S1~\,} & \textsf{\small 95.9}{\small{} } & \textsf{\small P8} &  & \textsf{\small 96.7}{\small{} } & \textsf{\small P8} &  & \textsf{\small 95.0}{\small{} } & \textsf{\small CP6} &  & \textsf{\small 95.9}{\small{} } & \textsf{\small P8}\tabularnewline
\textsf{\small S2~\,} & \textsf{\small 77.7}{\small{} } & \textsf{\small FC4} &  & \textsf{\small 78.5}{\small{} } & \textsf{\small C4} &  & \textsf{\small 83.5}{\small{} } & \textsf{\small P3} &  & \textsf{\small 81.8}{\small{} } & \textsf{\small P1}\tabularnewline
\textsf{\small S3~\,} & \textsf{\small 81.0}{\small{} } & \textsf{\small PO4} &  & \textsf{\small 81.0}{\small{} } & \textsf{\small CPz} &  & \textsf{\small 86.0}{\small{} } & \textsf{\small P2} &  & \textsf{\small 86.0}{\small{} } & \textsf{\small CP4}\tabularnewline
\textsf{\small S4~\,} & \textsf{\small 86.7}{\small{} } & \textsf{\small P7} &  & \textsf{\small 82.5}{\small{} } & \textsf{\small Pz} &  & \textsf{\small 89.2}{\small{} } & \textsf{\small P1,P2} &  & \textsf{\small 88.3}{\small{} } & \textsf{\small P4,PO4}\tabularnewline
\textsf{\small S5~\,} & \textsf{\small 86.0}{\small{} } & \textsf{\small P4} &  & \textsf{\small 92.6}{\small{} } & \textsf{\small P4} &  & \textsf{\small 88.4}{\small{} } & \textsf{\small CPz} &  & \textsf{\small 88.4}{\small{} } & \textsf{\small P4}\tabularnewline
\textsf{\small S6~\,} & \textsf{\small 68.3}{\small{} } & \textsf{\small P7,P3,PO3} &  & \textsf{\small 72.5}{\small{} } & \textsf{\small F7,FC3} &  & \textsf{\small 73.3}{\small{} } & \textsf{\small FT7} &  & \textsf{\small 74.2}{\small{} } & \textsf{\small Pz}\tabularnewline
\textsf{\small S7~\,} & \textsf{\small 79.3}{\small{} } & \textsf{\small PO3,O1} &  & \textsf{\small 86.8}{\small{} } & \textsf{\small PO3} &  & \textsf{\small 87.6}{\small{} } & \textsf{\small O1} &  & \textsf{\small 85.1}{\small{} } & \textsf{\small P3}\tabularnewline
\textsf{\small S8~\,} & \textsf{\small 87.6}{\small{} } & \textsf{\small O1} &  & \textsf{\small 81.0}{\small{} } & \textsf{\small C3,C6,O2} &  & \textsf{\small 90.9}{\small{} } & \textsf{\small P3} &  & \textsf{\small 91.7}{\small{} } & \textsf{\small PO3}\tabularnewline
\textsf{\small S9~\,} & \textsf{\small 77.7}{\small{} } & \textsf{\small C4} &  & \textsf{\small 76.0}{\small{} } & \textsf{\small TP9,PO3} &  & \textsf{\small 80.2}{\small{} } & \textsf{\small PO4} &  & \textsf{\small 80.2}{\small{} } & \textsf{\small PO4}\tabularnewline
\textsf{\small S10} & \textsf{\small 88.4}{\small{} } & \textsf{\small FC2,FC4} &  & \textsf{\small 86.8}{\small{} } & \textsf{\small P3} &  & \textsf{\small 93.4}{\small{} } & \textsf{\small CP3} &  & \textsf{\small 92.6}{\small{} } & \textsf{\small CP3}\tabularnewline
\textsf{\small S11} & \textsf{\small 82.6}{\small{} } & \textsf{\small FC2} &  & \textsf{\small 86.0}{\small{} } & \textsf{\small P7} &  & \textsf{\small 86.8}{\small{} } & \textsf{\small CP3} &  & \textsf{\small 86.8}{\small{} } & \textsf{\small CP5,CP3,P3}\tabularnewline
\hline 
\textsf{\small Avg.$\pm$std.} & \multicolumn{2}{l}{\textsf{\small 82.8$\pm$7.3}} &  & \multicolumn{2}{l}{\textsf{\small 83.7$\pm$7.1}} &  & \multicolumn{2}{l}{\textsf{\small 86.7$\pm$6.1}} &  & \multicolumn{2}{l}{\textsf{\small 86.4$\pm$6.1}}\tabularnewline
\hline 
\multicolumn{12}{l}{\textsf{\small Number of trials per participant: 121. Chance level:
50\%.}}\tabularnewline
\end{tabular}
\end{table}

\subsection*{Effect sizes}

We evaluated the practical significance of the results on Tables\,\ref{tab:Exp1a}-\ref{tab:ExpIII}
using effect sizes and 95\% confidence intervals. The results are
summarized in Table\,\ref{tab:Effect-sizes}. The tasks of classifying
the 8 initial consonants and 32 syllables rendered nearly the same
effect size. The classification of the 4 vowels, where the deviation
from the mean classification rate was relatively small resulted in
effect sizes equal to or greater than one standard deviation. An attentive reader may think that the results presented in Table\,\ref{tab:Effect-sizes} are at odds with those of Table\,\ref{tab:Exp1c}, where the surface Laplacian's mean classification rate was higher than that of the tangential field. We stress that such results are not inconsistent, as the effect size computation takes into account not only mean values, but also their variances. The
effect sizes were all positive in Exp.\,II, but the confidence interval
for the SL included zero, meaning that the hypothesis of no difference
in performance between the potential and SL could not be rejected
at 95\% level of confidence. In contrast, for this experiment the
superior performance of SL \& EF as compared to the potential was
confirmed with 2.1 standard deviations. The effect sizes were all
positive in Exp.\,III, but again the hypothesis of no difference
between the potential and SL could not be rejected because the confidence
interval included zero.

\begin{table}[H]
\textsf{\small \caption{\textsf{\small Effect sizes and 95\% confidence intervals for improvements
in classification rates.\label{tab:Effect-sizes}}}
}{\small \par}

\begin{tabular}{lr@{\extracolsep{0pt}.}lr@{\extracolsep{0pt}.}lr@{\extracolsep{0pt}.}l}
\hline 
\multicolumn{1}{c}{\textsf{\small Classification task}} & \multicolumn{2}{c}{\textsf{\small SL}} & \multicolumn{2}{c}{\textsf{\small EF}} & \multicolumn{2}{c}{\textsf{\small SL \& EF}}\tabularnewline
\hline 
\textsf{\small Exp.\,I (initial consonants)} & \textsf{\small 0}&\textsf{\small 6 (0.4 to 0.8)} & \textsf{\small 0}&\textsf{\small 9 (0.6 to 1.2)} & \textsf{\small 1}&\textsf{\small 0 (0.7 to 1.4)}{\small{} }\tabularnewline
\textsf{\small Exp.\,I (syllables)} & \textsf{\small 0}&\textsf{\small 6 (0.4 to 0.7)} & \textsf{\small 0}&\textsf{\small 7 (0.5 to 0.9)} & \textsf{\small 0}&\textsf{\small 8 (0.6 to1.0)}{\small{} }\tabularnewline
\textsf{\small Exp.\,I (vowels)} & \textsf{\small 1}&\textsf{\small 0 (0.6 to 1.3)} & \textsf{\small 1}&\textsf{\small 2 (0.6 to 1.7)} & \textsf{\small 1}&\textsf{\small 4 (1.0 to 1.9)}{\small{} }\tabularnewline
\textsf{\small Exp.\,II } & \textsf{\small 0}&\textsf{\small 1 ($-0.2$ to 0.3)} & \textsf{\small 1}&\textsf{\small 3 (0.4 to 2.1) } & \textsf{\small 2}&\textsf{\small 1 (0.9 to 3.4)}{\small{} }\tabularnewline
\textsf{\small Exp.\,III} & \textsf{\small 0}&\textsf{\small 1 ($-0.2$ to 0.5)} & \textsf{\small 0}&\textsf{\small 6 (0.3 to 0.9)} & \textsf{\small 0}&\textsf{\small 5 (0.3 to 0.8)}{\small{} }\tabularnewline
\hline 
\end{tabular}
\end{table}

\subsection*{Effect of smoothing}

We asked the question of whether improvements in classification rates
using SL, EF, or SL \& EF were a mere consequence of the $\lambda$
regularization, instead of reflecting an intrinsic capability of these
methods. If this hypothesis were true, then we should be able to improve
the performance of the electric potential by classifying its regularized
version. We tested this hypothesis by conducting another round of
classification in which the raw potential was repeatedly smoothed,
with $\lambda$ varying in the same log-scale grid from $\lambda=0.001$
to 100. We compared the results with the rates obtained with the raw
potential. The outcome of this analysis was that there was no significant
improvement in the classification rates due to the smoothing of the
potential. The largest effect size was 0.3 (95\%CI, 0.2--0.5) and
occurred in the classification of the 8 initial consonants. For all
other tasks the effect size was smaller than 0.3 and the confidence
interval included zero in all cases, supporting the null hypothesis
of no significant effect of regularization on the classification rates.

\subsection*{Classification with multiple channels}

We also evaluated the applicability of the electric field in multichannel
classification. In principle, the use of multiple channels permits
a fully exploitation of information content encoded in space and time,
meanwhile accounting for correlation between channels and inter-dependent
features that enlarge the number of false positives leading to misclassifications.
In order to perform this evaluation using LDA, we concatenated the
trials of the 5 and 10 best-performing channels disregarding their
physical locations. The enlarged signals were classified using the
same optimization procedure employed for single channels. Table\,\ref{tab:Multichannels}
shows the resulting effect sizes. The tables showing the classification
rates are presented in the Supplementary Material.

Small variations were observed in the performance of the SL, the most
important one occurring in the classification of the 8 initial consonants,
with a monotonic increase in effect size from 0.6 (Table\,\ref{tab:Effect-sizes})
to 0.7 and 0.8 (Table\,\ref{tab:Multichannels}). The effect of the
SL in the classification of the 4 vowels changed drastically, decreasing
10 folds for classification with the 10 best channels. The practical
effects of EF and SL \& EF were strongly affected in the classification
of the vowels, shape-color sensory images, and stop-go imagined images.
They remained about the same in the two other cases, except that the
effect of SL \& EF decreased from 1.0 (Table\,\ref{tab:Effect-sizes})
to 0.5 (Table\,\ref{tab:Multichannels}) in the classification of
the initial consonants using 10 channels. In comparison to the single-channel
classification, here the 95\% confidence interval included zero in
several cases, indicating loss of significance of improvements in
classification rate. 
\begin{table}[H]
\noindent \textsf{\caption{\textsf{Effect sizes for all classification tasks by concatenating
the 5 and 10 best-performing channels. The columns account for the
normal field, tangential field, and the total scalp electric field.\label{tab:Multichannels}}}
}

\begin{tabular}{lr@{\extracolsep{0pt}.}lr@{\extracolsep{0pt}.}lr@{\extracolsep{0pt}.}lcr@{\extracolsep{0pt}.}lr@{\extracolsep{0pt}.}lr@{\extracolsep{0pt}.}l}
\hline 
 & \multicolumn{6}{c}{\textsf{\small Five channels}} &  & \multicolumn{6}{c}{\textsf{\small Ten channels}}\tabularnewline
\cline{2-14} 
\multicolumn{1}{c}{\textsf{\small Classification task}} & \multicolumn{2}{c}{\textsf{\small SL}} & \multicolumn{2}{c}{\textsf{\small EF}} & \multicolumn{2}{c}{\textsf{\small SL \& EF}} &  & \multicolumn{2}{c}{\textsf{\small SL}} & \multicolumn{2}{c}{\textsf{\small EF}} & \multicolumn{2}{c}{\textsf{\small SL \& EF}}\tabularnewline
\hline 
\textsf{\small Exp.\,I (initial consonants)} & \textsf{\small 0}&\textsf{\small 7} & \textsf{\small 0}&\textsf{\small 8} & \textsf{\small 0}&\textsf{\small 8}{\small{} } &  & \textsf{\small 0}&\textsf{\small 8}{\small{} } & \textsf{\small 0}&\textsf{\small 8}{\small{} } & \textsf{\small 0}&\textsf{\small 5}{\small{} }\tabularnewline
\textsf{\small Exp.\,I (syllables)} & \textsf{\small 0}&\textsf{\small 7} & \textsf{\small 0}&\textsf{\small 6} & \textsf{\small 0}&\textsf{\small 8}{\small{} } &  & \textsf{\small 0}&\textsf{\small 6} & \textsf{\small 0}&\textsf{\small 8}{\small{} } & \textsf{\small 0}&\textsf{\small 6}{\small{} }\tabularnewline
\textsf{\small Exp.\,I (vowels)} & \textsf{\small 0}&\textsf{\small 4} & \textsf{\small 0}&\textsf{\small 1} & \textsf{\small $-$0}&\textsf{\small 1$^{\ast}$}{\small{} } &  & \textsf{\small 0}&\textsf{\small 1$^{\ast}$}{\small{} } & \textsf{\small $-0$}&\textsf{\small 4$^{\ast}$}{\small{} } & \textsf{\small $-0$}&\textsf{\small 9}{\small{} }\tabularnewline
\textsf{\small Exp.\,II} & \textsf{\small $-$0}&\textsf{\small 2$^{\ast}$} & \textsf{\small 0}&\textsf{\small 1} & \textsf{\small 0}&\textsf{\small 4}{\small{} } &  & \textsf{\small 0}&\textsf{\small 2$^{\ast}$}{\small{} } & \textsf{\small 0}&\textsf{\small 2$^{\ast}$}{\small{} } & \textsf{\small 0}&\textsf{\small 3$^{\ast}$}{\small{} }\tabularnewline
\textsf{\small Exp.\,III} & \textsf{\small 0}&\textsf{\small 0$^{\ast}$} & \textsf{\small $-$0}&\textsf{\small 2$^{\ast}$} & \textsf{\small $-$0}&\textsf{\small 4$^{\ast}$}{\small{} } &  & \textsf{\small 0}&\textsf{\small 1$^{\ast}$}{\small{} } & \textsf{\small 0}&\textsf{\small 0$^{\ast}$}{\small{} } & \textsf{\small $-$0}&\textsf{\small 3$^{\ast}$}{\small{} }\tabularnewline
\hline 
\multicolumn{14}{l}{\textsf{\small $^{\ast}$95\%CI includes zero.}}\tabularnewline
\end{tabular}
\end{table}

\section*{Discussion}

Overall the classification results were more accurate with the joint
use of the surface Laplacian of the potential and the tangential electric
field. The only exception was the mental task of Exp.\,III, for which
the electric field alone was more accurate in average than any other
waveform. However, this does not invalidate the view that the surface
Laplacian of the potential and the tangential electric field should
be used together to best assess non-overlapping information encoded
in different spatial directions. Nevertheless, this exception illustrates
a possible situation that may not be possible to predict, and that
can be associated to particular features of our experiments.

Improvements with SL \& EF were generally less significant in multichannel
classification. While the performance of the potential increased substantially
with multiple channels, the other waveforms were only slightly more
accurate, in some cases yielding effect sizes one standard deviation
or more smaller than obtained with using single channels, and in some
cases even negative effect sizes. A possible explanation for this
decline in performance may be related to the way we concatenate channels
for classification. Such concatenation, which has little practical
effect on the nonlocal electric potential disregards the electrodes'
physical locations, thus worsening the estimation of the surface Laplacian
differentiation and the electric field, which are local quantities.

We recall that the locality of the electric field and the surface
Laplacian was a primary reason for conducting this study. Presumably,
this property should reflect in a better identification of those brain
areas involved in the task performance, opening a prospect for applications
on EEG brain mapping and suggesting a criterion for a prior selection
of channels to perform classification. The three tasks studied in
Exp.\,I were important to ascertain this hypothesis, since they involved
auditory evoked activity and the signals were recorded with a high-density
electrode net. In good agreement with anatomical reports \citep[e.g.,][]{Hashimoto2000},
our results with SL \& EF showed the best performing channel being
close to A1 and A2 for all subjects, challenging the notion that the
EEG is an unreliable detector of localized activation due to poor
spatial resolution.

Recognizing vowels from syllables was the most challenging task encountered
in our study. The comparatively low rates achieved in this case were
possibly related to variations in the onset of vowels preceded by
different consonants. As long-duration consonants take longer to be
perceived as compared to short consonants, the onset of the ensuing
vowel varied affecting the classification negatively. Evidence of
variations in onset was reported, for instance, by \citet{Lawson1981}
based on the analysis of evoked potential.

Our study had several limitations that may have prevented a better
assessment of the true capability of the combined use of the surface
Laplacian and electric field to improve classification. The use of
a spherical scalp model was one of such limitations. For instance,
\citet{Babiloni1996,Babiloni1998} reported significant improvements
in surface Laplacian estimation by reconstructing the scalp surface
with magnetic resonance (MR). Also worthy of mention is the work of
\citet{He2001} in which the surface Laplacian was reliably estimated
using realistic electrode locations. It seems, therefore, plausible
to conjecture that a similar improvement could occur here, provided
that supplementary resources such as MR were available to perform
spatial differentiations more accurately. Finally, we remark that
numerical differentiations are inevitably affected by noise and the
mechanism of $\lambda$ regularization has a limit power to prevent
such effect. This limitation could be mitigated by the use of a more
sophisticated statistical technique for handling noise.

\section*{Conclusion}

This paper discussed the method of joint use of the surface Laplacian
differentiation and tangential electric field to improve EEG classification.
Its effectiveness was evaluated in the challenging problem of EEG
classification using a variety of experimental conditions. In all
experimental conditions (with one exception discussed above), the
joint use of the surface Laplacian and tangential electric fields
resulted in better classification rates for single electrode sites.
The statistical results were quite significant in most cases, supporting
a more extensive investigation of this approach in EEG analysis.

\begin{acknowledgements}
We would like to thank Dr. Gary Oas and Prof. Jos\'e Paulo de Mendon\c{c}a
for fruitful discussions on the electric field of the brain. We are
also in debt to Dr. Rui Wang for her help in all matters related to
Exp.\,I.\end{acknowledgements}

\appendix

\subsection*{Appendix}

This appendix provides an intuitive explanation for the relation between
the surface Laplacian of the scalp potential and the normal component
of the scalp electric field when recording the EEG. For clarity of
exposition, we focus on the current density $\mathbf{J}$, which is
a vector quantity that is locally related to the electric field $\mathbf{E}$
in the extracellular space by $\mathbf{J}=\sigma\,\mathbf{E}$, where
$\sigma$ is the electrical conductivity of the medium. The current
density is quasistatic continuous \citep[e.g.,][]{Plonsey1969,Haus1989},
thus obeying $\nabla\cdot\mathbf{J}=0$. Using spherical coordinates, this equation can be expressed
in the form 
\begin{equation}
\frac{1}{r^{2}}\frac{\partial}{\partial r}\left(r^{2}J_{r}\right)+\frac{1}{r\sin\theta}\frac{\partial}{\partial\theta}\left(\sin\theta\, J_{\theta}\right)+\frac{1}{r\,\sin\theta}\frac{\partial J_{\varphi}}{\partial\varphi}=0.
\end{equation}
We can conveniently rewrite this equation as 
\begin{equation}
\nabla_{s}\cdot\mathbf{J}_{s}=-\frac{2}{r}J_{r}-\frac{\partial J_{r}}{\partial r},\label{eq:The_surface_divergence_of_J}
\end{equation}
where $\nabla_{s}\cdot\mathbf{J}_{s}$ is the surface divergence of
the tangential current density $\mathbf{J}_{s}=J_{\theta}\hat{\bm{\theta}}+J_{\varphi}\hat{\bm{\varphi}}$.

The term $(2/r)\, J_{r}$ in the right-hand side of equation \eqref{eq:The_surface_divergence_of_J}
represents the contribution of the spherical geometry to lateral spread
of current. This term vanishes as $r$ goes to infinity, where the
sphere looks locally like a plane and the geometry does not affect
current flow in the normal direction. This limit corresponds to the
 model studied by \citet{He1992} and \citet{He1995}. The
term $\partial J_{r}/\partial r$ represents the rate of vanishing
of $J_{r}$ in the normal direction, and is particularly significant
at the scalp-air interface, where the negligibility of the air conductivity
causes the abrupt vanishing of $J_{r}$ along the outer scalp surface.
Hence, an intuitive interpretation of equation \eqref{eq:The_surface_divergence_of_J}
is that it describes how the geometry and changes in flux in the normal
direction affect the behavior of $\mathbf{J}_{s}$, so as to ensure
the continuity of the total current density.

Both terms on the right-hand side of equation \eqref{eq:The_surface_divergence_of_J}
depend on boundary conditions. Assuming that the scalp is surrounded
by air implies the vanishing of $J_{r}$ along the outer scalp surface
due to the negligibility of the air conductivity, which is about 14
orders of magnitude smaller than the scalp conductivity and prevents
current to exit the head through the scalp-air interface. The recording
of EEG signal changes this condition locally, causing, inevitably,
an outflow of current beneath the measurement electrodes. Presumably,
without the constraint of the abrupt vanishing of $J_{r}$ the magnitude
of $\partial J_{r}/\partial r$ becomes smaller at these locations,
so that, at least to the lowest order of approximation, we can use \eqref{eq:The_surface_divergence_of_J} to write 
\begin{equation}
\nabla_{s}\cdot\mathbf{J}_{s}(\mathbf{r})\approx-\frac{2}{r_{\mathrm{scalp}}}J_{r}(\mathbf{r}),\label{eq:div_Js}
\end{equation}
it being understood that the position $\mathbf{r}$ coincides 
with a scalp electrode location.

Let $\sigma_{s}^{\mathrm{scalp}}$ and $\sigma_{r}^{\mathrm{scalp}}$
represent the tangential conductivity and the radial conductivity
along the scalp. Substituting $J_{r}=\sigma_{r}^{\mathrm{scalp}}E_{r}^{\mathrm{scalp}}$
and $\mathbf{J}_{s}=\sigma_{s}^{\mathrm{scalp}}\,\mathbf{E}_{s}^{\mathrm{scalp}}=-\sigma_{s}\nabla_{s}\Phi_{s}^{\mathrm{scalp}}$,
where $\nabla_{s}\Phi_{s}^{\mathrm{scalp}}$ is the surface gradient
of the surface potential $\Phi_{s}^{\mathrm{scalp}}$, and assuming
that the conductivities $\sigma_{r}^{\mathrm{scalp}}$ and $\sigma_{s}^{\mathrm{scalp}}$
are approximately constants, we obtain from \eqref{eq:div_Js} 
\begin{equation}
E_{r}^{\mathrm{scalp}}(\mathbf{r})\approx\frac{r_{\mathrm{scalp}}}{2}\frac{\sigma_{s}^{\mathrm{scalp}}}{\sigma_{r}^{\mathrm{scalp}}}\nabla_{s}^{2}\Phi_{s}^{\mathrm{scalp}}(\mathbf{r}).\label{eq:Approximation_for_Er}
\end{equation}
Since $E_{r}^{\mathrm{scalp}}$ is locally related to $J_{r}^{\mathrm{scalp}}$,
this expression agrees with the usual view that the surface Laplacian
differentiation provides a good method to associate local EEG events
generated by cortical radial dipoles to the underlying physical structure.
But the approximation \eqref{eq:Approximation_for_Er} was obtained
without any assumption about brain sources.

The scalp tangential conductivity $\sigma_{s}^{\mathrm{scalp}}$ and
the radial conductivity $\sigma_{r}^{\mathrm{scalp}}$ were introduced
to account for a prominent directional dependency in the scalp structure,
as pointed out by experimental studies \citep[e.g.,][]{Abascal2008,Petrov2012}.
Accounting for this anisotropy requires a tensor representation for
$\sigma^{\mathrm{scalp}}$, which in our model was written 
\begin{equation}
\bm{\sigma}^{\mathrm{scalp}}=\sigma_{r}^{\mathrm{scalp}}\hat{\mathbf{r}}\hat{\mathbf{r}}+\sigma_{s}^{\mathrm{scalp}}\hat{\bm{\theta}}\hat{\bm{\theta}}+\sigma_{s}^{\mathrm{scalp}}\hat{\bm{\varphi}}\hat{\bm{\varphi}}.\label{eq:scalp-tensor-conductivity}
\end{equation}
Typically, the ratio $\sigma_{s}^{\mathrm{scalp}}/\sigma_{r}^{\mathrm{scalp}}$
is about 1.5, so that the multiplying factor $r_{\mathrm{scalp}}\sigma_{s}^{\mathrm{scalp}}/2\sigma_{r}^{\mathrm{scalp}}$
in \eqref{eq:Approximation_for_Er} is approximately 7.0\,cm for
typical values of the head radius.
\end{document}